# The Neural Networks Based Needle Detection for Medical Retinal Surgery

Jidong Xu[1,*], Jinglun Yu[1,*], Jianing Yao[2], and Rendong Zhang[1]

[1]Department of Electrical and Computer Engineering, Johns Hopkins University, Baltimore, United States.

[2]Department of Biostatistics, Johns Hopkins Bloomberg School of Public Health, Baltimore, United States.

*These authors contributed equally to the work and should be regard as co-first author.

Jidong Xu: Email: jxu121@jhu.edu, Jinglun Yu: Email: jyu146@jhu.edu, Jianing Yao: Email: jyao37@jhmi.edu, Rendong Zhang: Email: rzhang86@jhu.edu.

**Abstract**

In recent years, deep learning technology has developed rapidly, and the application of deep neural networks in the medical image processing field has become the focus of the spotlight. This paper aims to achieve needle position detection in medical retinal surgery by adopting the target detection algorithm based on YOLOv5 as the basic deep neural network model. The state-of-the-art needle detection approaches for medical surgery mainly focus on needle structure segmentation. Instead of the needle segmentation, the proposed method in this paper contains the angle examination during the needle detection process. This approach also adopts a novel classification method based on the different positions of the needle to improve the model. The experiments demonstrate that the proposed network can accurately detect the needle position and measure the needle angle. The performance test of the proposed method achieves 4.80 for the average Euclidean distance between the detected tip position and the actual tip position. It also obtains an average error of 0.85 degrees for the tip angle across all test sets.

**Keywords**-YOLOv5, deep neural networks, classification method, needle detection, retinal surgery

## 1. Introduction

Nowadays, needles are common surgical devices in retinal surgery. The insertion angle and position of the needle are critical to the surgery procedure. In actual practice, it is often difficult to obtain a quick and accurate observation of the angle and tips of the tiny needle. Meanwhile, owing to the variable angles and tips direction that can adopt, the detection of needles becomes a relatively challenging task. Therefore, the application of deep neural networks is necessary. The convolutional neural network (CNN) is widely used in image processing due to its powerful feature extraction capability, particularly in object detection [1]. Many classical algorithms are applied to achieve a fast and accurate detection of objects [2]. Such as R-CNN based on SVM is a classical deep learning method to do object detection [3][4][5], Fast R-CNN [6] based on VGG-16 [7], and the Faster R-CNN [8] based on Region Proposal Network (RPN). However, compared with the most advanced Faster-RCNN among the state-of-the-art models, the proposed model is based on YOLOv5, which introduces Focus Layer in the backbone, and achieves a higher speed performance and superb accuracy for detection [9]. For the needle position detection in the surgery process, the first need is a robust classification method that allows the needle form to be discriminated. The proposed approach develops the angle, and tip sets for needle detection, indicating that the intuitive and specific classification with refined deep neural networks provides exemplary performance in the detection process [10][11]. The proposed classification method separates all angle and tip conditions simply and intuitively into four classes. This process reduces the time consumed by training and simplifies the tedious training process while ensuring detection accuracy. Application to the pig retina background data set revealed that the refined YOLOv5 model and the classification method in the experiments gives superior results in the detection accuracy.

## 2. Approach

In this part, we discuss the specific method to determine the needle position on medical retinal surgery. Our model is based on YOLOv5 object detection model, which is a state-of-the-art neural network framework. Furthermore, we propose a novel classification approach achieving the accurate detection of the needle tips' position and needle angles.

## 2.1. YOLO Framework

You Only Look Once (YOLOv5) is a CNN-based neural network [9][12][13], which is able to achieve a fast and accurate object detection. The input of YOLOv5 model is a whole image, and the output of YOLOv5 are the predicted bounding boxes' coordinates, object classes and the corresponding confidence of the prediction. In YOLOv5 model, the input image is divided into S*S grid cells. Next, YOLOv5 model applies object localization and classification on each grid cells and finds the grid cells which includes the centre point of the object bounding box. After that, applying Intersection over Union [14] and Non-Maximum Suppression [15] method, the YOLO framework predicts the final output, which is the bounding boxes containing the predicted labels, probabilities, and most importantly, the coordinates of the bounding boxes.

## 2.2. Detection of Needle Tips and Angles

After making a prediction on the input image, the YOLOv5 model will output a bounding box, which contains the box vertex coordinates, object classes and corresponding probabilities. Since we would like to determine the needle tips' position and needle angles, we make the needle tip locate on one of the vertexes of the bounding boxes when we label the object before training. Besides, the middle point of the needle is placed on the corresponding diagonal vertex. Therefore, if we are able to determine the coordinates of vertexes of the bounding boxes and determine the needle tip is located at which vertex of the bounding box, we can detect the needle tips position. As for the needle angles, we calculate the angle between the needle and the horizontal edge of the bounding box as the needle angle. Therefore, we could calculate the needle angle with this equation:

$$\tan \theta = \frac{|y_1 - y_2|}{|x_1 - x_2|} \quad (1)$$

where $(x_1, y_1)$ is the coordinate of the output predicted needle tip, and the $(x_2, y_2)$ is the predicted coordinate of the midpoint of the needle. With these two coordinates, we are able to determine the needle angles.

As we know, the needle tip is located at one of the vertex of the predicted bounding boxes, and there are four different possible situations of the needle tip's position: the needle tip may be located at the left top vertex of the bounding box (LT), on the left bottom vertex of the bounding box (LB), on the right top vertex (RT) or on the right bottom vertex (RB), as shown in figure 1. In order to determine the needle tips' locating on which vertex and needle angles, we propose a method, which divides the needle images into four classes according to their needle tips locations in the train set. The needles with tips locating on the left top vertex of the bounding box is classified as one class and is labelled as LT. The second class is the needles with the tips locating on the left bottom vertex of the bounding box, and is labelled as LB. Similarly, we classify the other two needle classes called RT (right-top) and RB (right-bottom). In this way, the output of YOLOv5 model includes the needle classes and the bounding boxes' coordinates. With the classification of needle tips, we are able to determine the needle tip's position and finally calculate the needle angle.

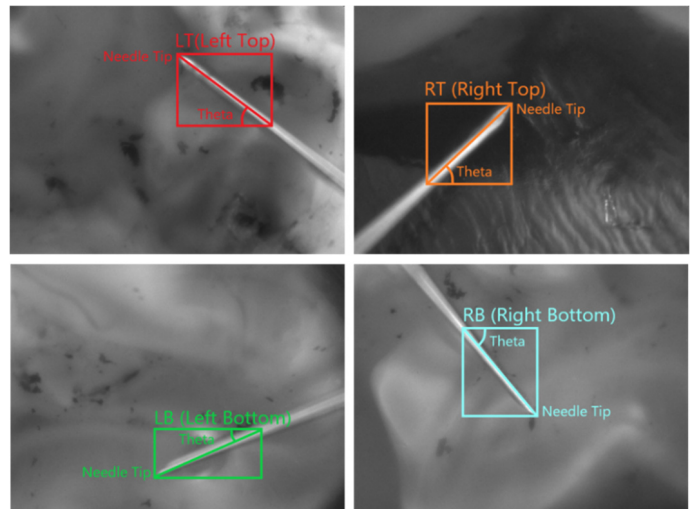

Figure 1. Four classes of needles: LT, LB, RT, RB.

## 3. Experiment results

In the experiment, we collect 120 needle images as the train set and test set, which are all shoot by microscope. The diameter of the needle is 0.4 millimetre. Since there is a high degree of similarity between pig retinal images and human retinal images, we apple the pig retinas as the background to collect the original needle image in random positions over the background. The image size of every original image data is 692x516.

The original needle images are divided into two parts, which includes the train set and the test set. The train set contains 96 needle images, and the test set contains 24 images. For the train set, we apply data augmentation on the original 96 images and enlarge the number of the training images into 576. The data augmentation includes image flip, rotation, and corruption. In train set, we label each needle image in YOLOv5 format. Each needle image is assigned a class label (LT, LB, RT or RB) and a bounding box indicating the needle position. For the training, we set the batch size as 4, set the training epochs as 200, and use the pre-trained weights start the training.

### 3.1. Training results

The training results are shown in figure 2. With the training epochs close to 200, the box loss, object loss and classification loss decrease and close to 0. The box loss is used to calculate position loss between the predicted bounding boxes and the ground truth bounding boxes. The object loss calculates the loss of between the predicted probability and ground truth about whether there is an object needle existing in the image. And the class loss calculates the difference of the probability between the predicted class and ground truth class. The precision and recall rate are close to 1 at 200 epochs, which indicates this object detector can be said as a good model. Furthermore, the mAP@.5:.95 is more than 0.9. These results show the model training satisfies the experiment requirement.

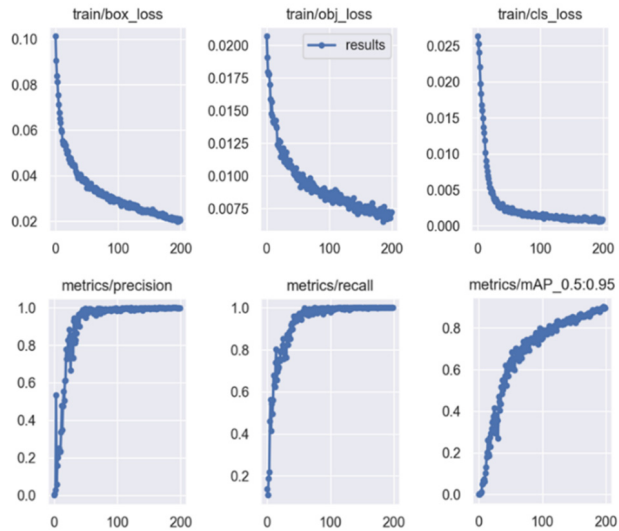

Figure 2. Training results.

### 3.2. Testing results

After training the model, we apply our trained model on the test set, which contains 24 needle images with different needle positions and needle angles. The detection results are shown in figure 3. In every test image, there is a detected object, which includes the bounding box, predicted class label and the corresponding confidence. From the test results, we compare all the predicted class labels with the ground truth labels, and the results show all the detected needles have corrected predicted labels. Besides, with observation, the predicted bounding boxes are accurate since the needle positions and directions are close to the ground truth.

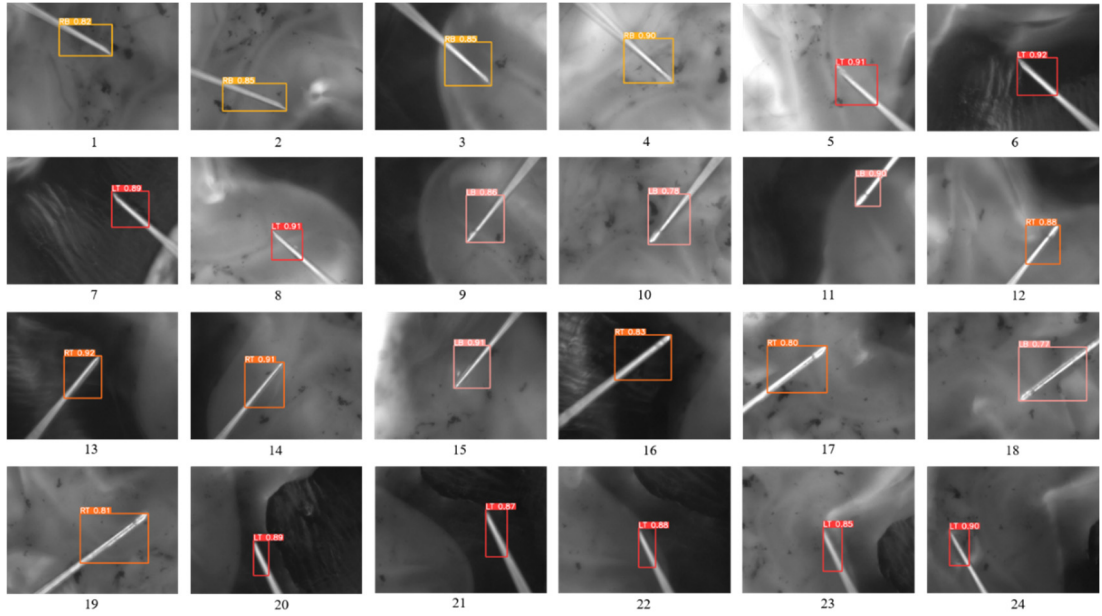

Figure 3. Detecting results.

The detected needle tips' positions and the detected needle angles are shown in Table 1. The Det Tip is the detected needle tip's coordinates in pixel, and the Real Tip is the ground truth coordinate of needle tip's position; the Det Ang is the detected needle angle, and the Real Ang is the ground truth needle angle; the Tip Dist. shows the Euclidian distance between the detected needle tip's position and the ground truth needle tip's position; the Ang Err shows the error of the detected needle angle and the ground truth needle angle. In order to evaluate the average detecting performance of the model on all test images, we calculate the average Euclidian distance between all the predicted needle tips and ground truth needle tips, which is 4.8 in pixel. Besides, we also calculate the average angle error for all test images, which is 0.85 degree. Furthermore, the detection time on each test image is about 0.015s. These test results indicate our model achieves an accurate and fast needle detection.

Table 1. Needle tips' positions and needle angles.

| Image | Det Tip | Real Tip | Det Ang | Real Ang | Tip Dist | Ang Err |
|---|---|---|---|---|---|---|
| 1 | (424,215) | (422,212) | 30.81 | 31.04 | 3.61 | 0.23 |
| 2 | (385,438) | (388,436) | 23.44 | 23.59 | 3.61 | 0.15 |
| 3 | (469,336) | (469,338) | 43.27 | 42.73 | 2.00 | 0.54 |
| 4 | (460,325) | (468,335) | 42.13 | 42.66 | 12.81 | 0.53 |
| 5 | (372,247) | (372,247) | 44.13 | 43.79 | 0.00 | 0.34 |
| 6 | (362,218) | (361,215) | 43.35 | 44.04 | 3.16 | 0.69 |
| 7 | (424,138) | (426,138) | 44.23 | 44.19 | 2.00 | 0.04 |
| 8 | (326,292) | (329,293) | 45.00 | 44.07 | 3.16 | 0.93 |
| 9 | (367,345) | (364,346) | 51.30 | 51.04 | 3.16 | 0.26 |
| 10 | (358,354) | (364,345) | 50.66 | 51.53 | 10.82 | 0.87 |
| 11 | (451,199) | (456,195) | 50.43 | 50.24 | 6.40 | 0.19 |
| 12 | (534,275) | (531,276) | 49.07 | 50.89 | 3.16 | 1.82 |
| 13 | (383,177) | (379,178) | 48.55 | 49.24 | 4.12 | 0.69 |
| 14 | (377,203) | (376,204) | 49.19 | 50.99 | 1.41 | 1.80 |
| 15 | (319,308) | (317,311) | 50.03 | 51.13 | 3.61 | 1.10 |
| 16 | (453,92) | (452,91) | 38.87 | 38.34 | 1.41 | 0.53 |
| 17 | (337,136) | (337,139) | 38.37 | 37.69 | 3.00 | 0.68 |
| 18 | (367,359) | (371,361) | 38.51 | 37.99 | 4.47 | 0.52 |
| 19 | (572,190) | (570,187) | 36.16 | 37.20 | 3.61 | 1.04 |
| 20 | (254,299) | (257,298) | 66.75 | 68.03 | 3.16 | 1.28 |
| 21 | (446,174) | (449,171) | 65.98 | 67.28 | 4.24 | 1.30 |
| 22 | (322,250) | (326,251) | 66.97 | 67.41 | 4.12 | 0.44 |
| 23 | (324,246) | (326,223) | 67.11 | 70.89 | 23.09 | 3.78 |
| 24 | (90,254) | (86,251) | 62.07 | 61.39 | 5.00 | 0.68 |

## 4. Conclusions

In this paper, we present a new model for needle position detection for medical retinal surgery use. The proposed approach based on the YOLOv5 object detection algorithm provides a novel classification method for detecting different positions and angles of needle tips. The experiment results demonstrate that this model can quickly and accurately extract the figure of needle tips and corresponding angles with a small error. Furthermore, applying the model in ex-vivo pig retinal data sets obtains superior performance in the needle detection tasks. In the following development, we will march forward in two aspects. First, we will apply data in real surgery or use the human retina as background for more authentic approaches. Second, to improve the proposed method's applicability, we will enlarge the dataset with multiple needle conditions and apply needle detection in real-time retinal surgery based on video instead of image [16].

## Acknowledgments

Here, we would like to express our thanks to all the people who provide us assistance and encouragement in this paper. Firstly, we want to thank every member of our team. With the efforts and persistence of every group member, we overcome difficult and challenges and complete our paper. Furthermore, we would like to express our acknowledgments to our families and friends, especially to our parents, who provide us endless love and unwavering support.